\documentclass{elsart}

\usepackage{natbib}
\usepackage{graphicx}

\usepackage{amssymb}

\def\simlt{\mathrel{\hbox{\rlap{\hbox{\lower4pt\hbox{$\sim$}}}\hbox{$<$}}}}
\def\simgt{\mathrel{\hbox{\rlap{\hbox{\lower4pt\hbox{$\sim$}}}\hbox{$>$}}}}
\newcommand{\lsim }{{\lower0.8ex\hbox{$\buildrel <\over\sim$}}}
\newcommand{\gsim }{{\lower0.8ex\hbox{$\buildrel >\over\sim$}}}
\newcommand{\lcgs}{\ifmmode erg~~s^{-1}\else erg~s$^{-1}$\fi}
\newcommand{\fcgs}{\ifmmode {\rm erg~cm}^{-2}~{\rm s}^{-1}\else
erg~cm$^{-2}$~s$^{-1}$\fi} 
\newcommand{\nh}{{\ifmmode{\rm N_{H}} \else N$_{H}$\fi}~}

\def\about{$\sim$}
\def\arcsec{$\,^{\prime\prime}$~}
\def\arcmin{$\,^\prime$}

\def\erg/cm2sec{ergs~cm$^{-2}$~s$^{-1}$}  
\def\ergcm2{ergs~cm$^{-2}$}

\def\X{$\times$}
\def\Fx{F$_X$~}
\def\Fv{F$_V$~}

\def\FxFv{{F$_x$/{F$_V$}}~}

\def\Halpha{H$\alpha$~}
\def\Lx{L$_x$~}

\def\pc3{pc$^{-3}$~}
\def\cm3{cm$^{-3}$~}

\begin{document}

\begin{frontmatter}



\title{Compact X-ray Binaries in and out of Core Collapsed Globulars}


\author{Jonathan E. Grindlay}

\address{Harvard-Smithsonian Center for Astrophysics, 60 Garden St.,
  Cambridge, MA 02138, USA}

\begin{abstract}
We review new Chandra and HST observations of the core collapsed 
cluster NGC 6397 as a guide to understanding the compact binary 
(CB) populations in core collapse globulars. New  
cataclysmic variables (CVs) and main sequence chromospherically 
active binaries (ABs) have been identified, enabling a larger 
sample for comparison of the \Lx, \FxFv and X-ray vs. optical color  
distributions. Comparison of the  numbers of CBs with 
\Lx \gsim10$^{31}$ \lcgs~ in 4 core collapse vs. 12 King model 
clusters reveals that the specific frequency S$_X$ (number of CBs 
per unit cluster mass) is enhanced in core collapse clusters, even 
when normalized for their stellar encounter rate. Although 
core collapse is halted by the dynamical heating due to stellar 
(and binary) interaction with CBs in the core, 
we conclude that production of the 
hardest CBs -- especially CVs -- is enhanced during core collapse.
NGC 6397 has its most luminous CVs nearest the cluster center, 
with two newly discovered very low luminosity (old, quiescent) 
CVs far from the core. The active binaries as well as neutron 
star systems (MSP and qLMXB) surround the central core. The 
overall CB population appears to be asymmetric about the cluster 
center, as in several other core collapse clusters 
observed with Chandra, suggesting still poorly-understood 
scattering processes.
\end{abstract}

\begin{keyword}
globular clusters \sep X-rays: general \sep binaries: close \sep 
cataclysmic variables \sep X-ray binaries \sep millisecond pulsars

\PACS 97.10.Gz \sep 97.30.Qt \sep 97.80.Gm \sep 97.80.Jp

\end{keyword}

\end{frontmatter}


\section{Introduction}
The high resolution imaging made possible by Chandra has enabled 
new insights into the compact binary population of globular cluster 
cores. In dense cores of globulars, compact or ``hard'' binaries with 
component stellar velocities much larger than the cluster velocity dispersion 
are both produced and exchanged. They also provide the internal energy 
source to reverse core collapse, the process whereby mass segregation 
would otherwise lead to stellar mergers and production of 
an intermediate mass black hole (IMBH) with a signature cusp in the 
central density and velocity dispersion of the cluster \citep{Bahcall76}. 
Chandra's \about0.5\arcsec angular resolution 
within its central $\sim$4\arcmin~ 
field of view is both sufficient to locate the compact binaries (CBs) 
and conduct the first X-ray studies of their spectra and time variability 
down to the \about10$^{29-32}$ \lcgs~ luminosities characteristic of 
the 4 major classes of CBs. These are: active binaries (ABs), or 
near-contact binaries of main sequence stars (e.g. BY Dra type systems) 
in which X-ray emission occurs by chromospheric-coronal activity of 
these rapidly rotating stars; accreting white dwarfs (WDs) with main 
sequence companions, detected as cataclysmic 
variables (CVs); quiescent low mass 
X-ray binaries (qLMXBs) in which a neutron star (NS) primary sporadically 
accretes from a main sequence companion (usually evolved); and millisecond 
pulsars (MSPs), in which the detached secondary of a (former) qLMXB has 
allowed the spun-up NS to emit X-rays by (predominantly) the thermal 
emission from its polar caps heated by positron return 
currents. All of these classes of CBs have been discovered and 
detected with Chandra in both relaxed King model clusters (e.g., 47
Tuc; see \citet{Grindlay01a} and \citet{Heinke05}) and 
post core collapse (PCC) clusters (e.g. 
NGC 6397; see \citet{Grindlay01b} and Grindlay et al (2006; 
hereafter GvBB06), in preparation), 
and a comparison of these two clusters has been given recently 
by \citet{Grindlay05}. 

Here we extend this comparison between PCC and King model 
clusters and present additional new Chandra results for the PCC  
cluster NGC 6397 to derive constraints on the CB population 
that may be attributable to the  core collapse process itself. 
We provide evidence for both CB destruction and creation before and 
during the PCC phase. Although only 2-3 of the MSPs in 47Tuc are 
likely doubly-exchanged, in NGC 6397 the single confirmed (radio) MSP 
as well as a second possible MSP (very similar in its X-ray and 
optical properties, but not detected as a radio MSP) has 
non thermal emission likely from shocked gas from its newly acquired 
main sequence companion, which is nearly filling 
its Roche lobe and interacting with the 
pulsar wind \citep{Grindlay02,Bogdanov05}. We present new results 
from both the Chandra (GvBB06) 
and HST (Cohn et al 2006, in preparation; hereafter CLC06) 
observations of NGC 6397, which have now allowed X-ray/optical 
identifications for most of the CBs in the cluster with 
\Lx \gsim10$^{29.5}$ \lcgs and have increased the number of 
confirmed CVs to 12 and possibly 13.
  
We then turn to other PCC clusters observed with Chandra for 
comparison. We find that PCC clusters have a higher specific 
frequency of CBs (mixture of all types, particularly 
CVs) than King model clusters. 
Additional very deep observations of NGC 6397 could provide important 
new tests of dynamical processes and binary populations in PCC clusters.

\section{New Results for NGC 6397}

\begin{figure}
\center{\includegraphics[width=14cm]{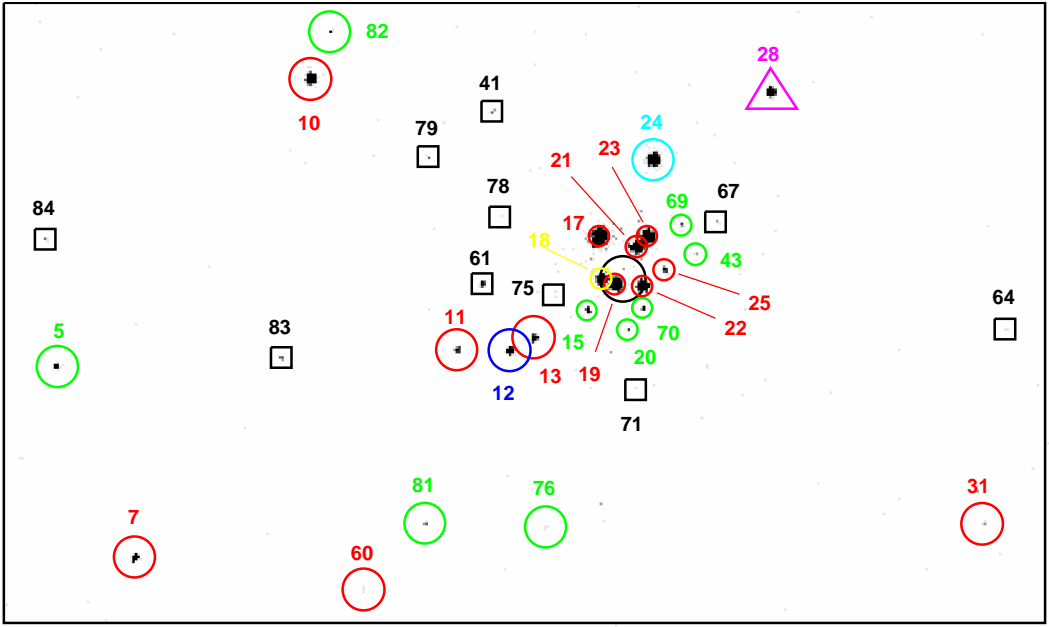}}
\caption{Merged ACIS-I (2000) and ACIS-S (2002) Chandra photon images 
(0.3 - 6.0 keV) of NGC 6397 for the 3.3\arcmin \X 2.0\arcmin~ region 
containing all 12 CVs identified thus far. The cluster center and  
optical core radius are measured (Taylor et al 2006, submitted; 
hereafter TGE06) 
to be 1\arcsec NW 
of source U19 (CV2) and 4.4\arcsec, respectively, as marked by the black 
circle. All sources optically identified as cluster members  
are circled and Chandra source U numbers 
(\citet{Grindlay01b}, GvBB06)   
are given: CVs (red), ABs (green), MSP (blue), candidate MSP (yellow), 
and the qLMXB (cyan). The qLMXB (U24) is not identified optically 
but is still circled as a certain cluster 
member. U28 (magenta triangle) is a background AGN. 
The 9 remaining unidentified sources in this region are marked by 
black squares and source numbers. Most are likely cluster 
sources, bringing the total number in 
this region to 32. Several single-pixel count 
excesses (e.g. to right of label for U76) are likely CR 
background events and not consistent with the psf of real 
sources. The cluster half-mass radius, r$_h$ = 2.3\arcmin, is just beyond 
the lower left corner of the box. Three AB counterparts are outside 
the box but inside r$_h$: U42 (north) and U74, U77 (south). An 
additional 7 unidentified sources (north) and 3 (south) are outside 
the box but inside r$_h$ and are probably dominated by the \about5-10 
background AGN expected inside r$_h$.} 
\label{fig:merge-I-S-ACS}
\end{figure}

At 2.3 kpc, NGC 6397 is the closest PCC globular cluster 
and only \about0.1kpc farther\citep{Harris96} than M4, the closest 
globular of all. Combined with its relatively low mass, 
its population of stars vs. compact binaries in its central 
core is (by far) the best-resolved of any PCC cluster. 
Despite its very high central density in its collapsed core, with 
core radius 4.4$\pm3.2$\arcsec for its brightest stars (V = 16.5-18.0) 
near the main sequence turnoff (TGE06), both HST imaging 
(\citet{Cool02}, TGE06, CLC06), and the first (July 2000) Chandra 
observations\citep{Grindlay01b} resolve the core completely. In 
\citet{Grindlay05} we presented initial results from the second 
(May 2002) Chandra observations made with ACIS-S (vs. ACIS-I for the 
first observation). Here we give additional results 
of these observations (see GvBB06 for a complete 
report) as well as our 
new HST/ACS observations (CLC06), which together provide 
a nearly completely identified population of CVs, ABs and compact 
binary X-ray sources in the cluster. Figure \ref{fig:merge-I-S-ACS} 
shows the central \about3.3\arcmin \X~ 2.0\arcmin~ of the cluster 
for the combined (merged) ACIS-I and ACIS-S data in the 0.3 - 6 keV 
band. 

\begin{figure}
\begin{center}
{\includegraphics[width=6.9cm]{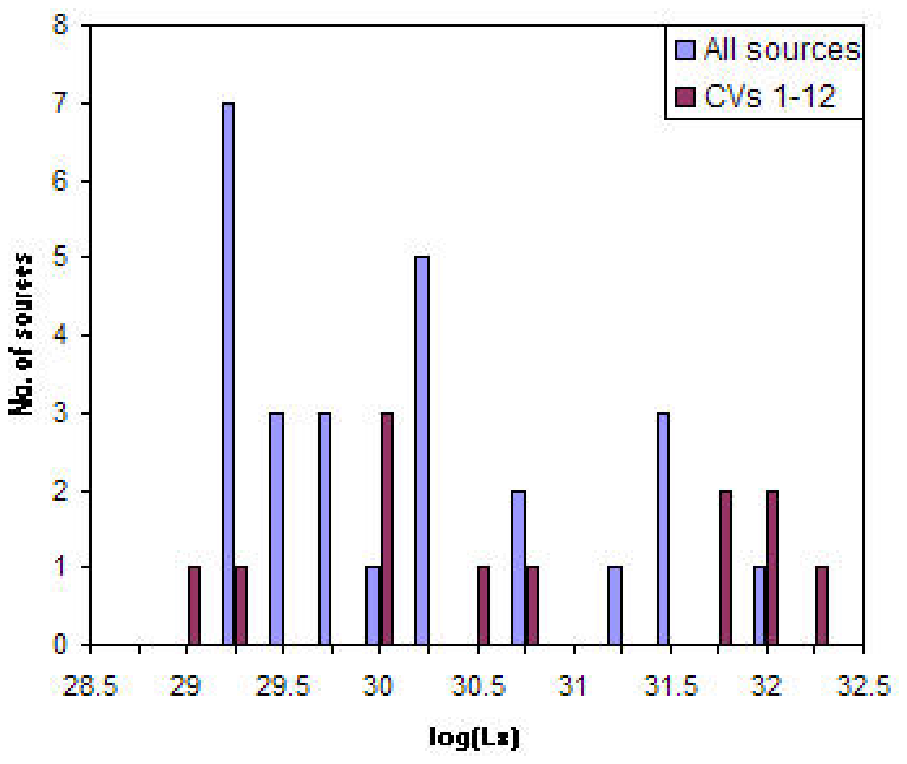}}
{\includegraphics[width=6.5cm,height=5.77cm]{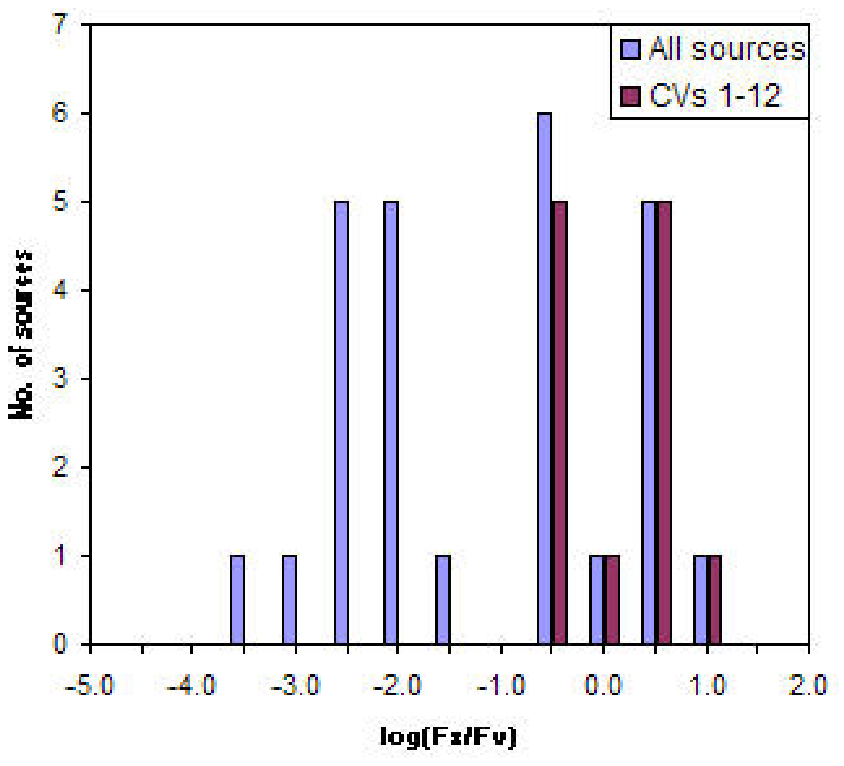}}
\end{center}
\caption{Distributions in \Lx (left) and \FxFv (right) for the 26 
sources in NGC 6397 with optical identifications and including 3 sources 
which are outside the box shown in Figure \ref{fig:merge-I-S-ACS}:  
U14, U42 and U77, which we identify with variables V35, V26 and V36 
\citep{Kaluzny06}, respectively. \Lx and \Fx are in the Sc 
band, and unabsorbed for the cluster log(\nh) = 21.0, as are 
the measured (V magnitudes) optical fluxes, \Fv. The 12 
CVs are compared with the full population to show their characteristically 
different distribution in \FxFv but not \Lx.}
\label{fig:Lx-FxFv-1keV}
\end{figure} 

The stacked ACIS-S (2 \X~ 28ksec exposures, May 13 \& 15, 2002) and ACIS-I 
(single 49ksec exposure, July 31, 2000) combined reach a 
detection threshold (3-5 cts) 
luminosity of \Lx \about 2 \X~ 10$^{29}$ \lcgs 
in the Sc band (0.5 - 2.0 keV) 
for an assumed characteristic source temperature of kT = 1 keV.  
This is a factor of 2 below the \Lx limit for the 
ACIS-I data alone, as reported by 
\citet{Grindlay01b}. For an assumed source spectrum with kT = 1 keV, 
as appropriate for the  optically identified ABs originally 
reported \citep{Grindlay01b} as well as the new AB 
counterparts identified by \citet{Kaluzny06} and CLC06, 
we show  in Figure \ref{fig:Lx-FxFv-1keV} the distributions 
in \Lx and \FxFv for the 23 sources in Figure \ref{fig:merge-I-S-ACS} 
with optical identifications plus 3 source optically identified 
just outside the box (see caption for Figure \ref{fig:Lx-FxFv-1keV}). 
Using a spectral temperature kT = 10
keV, as is appropriate for the 6-8 brightest CVs (\citet{Grindlay01b}, 
GvBB06), would give \Lx and \FxFv values larger by a factor 
of \about2. 

\begin{figure}
\begin{center}
{\includegraphics[width=12.cm]{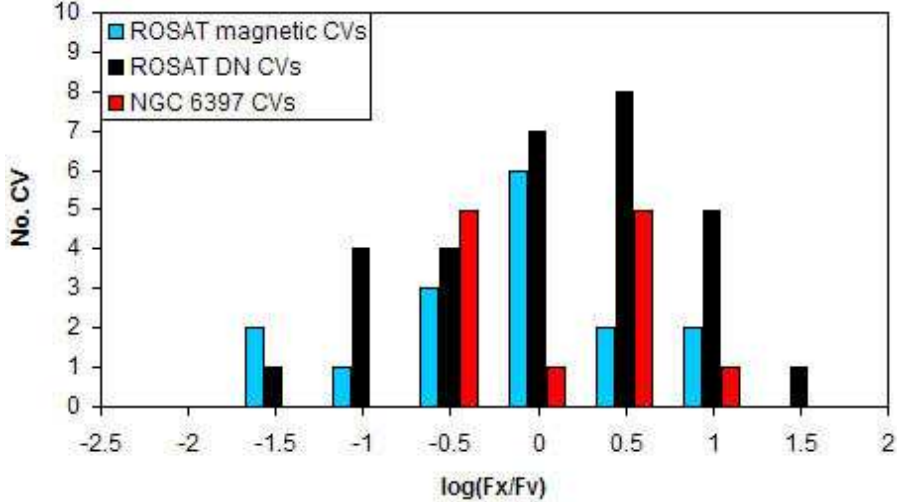}}
\end{center}
\caption{\FxFv distributions of ROSAT CVs (\Fx in ROSAT band, or 
approximately Sc band) in the local field for  
both magnetic CVs and dwarf novae
(from \citet{Verbunt97}), and the distribution for the 12 
CVs in NGC 6397 (unabsorbed Sc band), which are likely a mixture.}
\label{fig:CV-types-FxFv}
\end{figure}

It is striking how the 12 CVs (red histograms) are distinguished 
from the total population by having their log(\FxFv) values 
significantly larger than the full source population. This result is 
independent of the assumed X-ray spectrum: the \Lx and \Fx 
values in Figure \ref{fig:Lx-FxFv-1keV} are derived for an 
assumed Bremsstrahlung spectrum with kT = 1 keV, which fits 
most of the ABs (\citet{Grindlay01b}, GvBB06), whereas for 
the Hc band (2.0 - 8.0 keV) and a harder source spectrum with 
kT = 10 keV (appropriate to the bright CVs), the overall 
distributions are very similar but shifted to larger values 
by \about0.3 in log(\Lx) or in log(\FxFv) for the CVs.
The \FxFv distribution is consistent with that found for nearby 
field CVs observed with ROSAT \citep{Verbunt97}. Comparison 
of the \FxFv distribution (Figure \ref{fig:CV-types-FxFv}) 
for the NGC 6397 CVs with those for dwarf nova (DN) or magnetic 
(AM Her or DQ Her) CVs in the galactic disk observed with ROSAT  
shows the CVs in NGC 6397 are likely a mixture of both types. 
Indeed, recent HST studies \citep{Shara05} show that at least 
two of the luminous CVs near the cluster center undergo 
dwarf nova (DN) like outbursts. While these may still be magnetic 
CVs (DQ Her type), as originally suggested on the basis of 
their HeII emission \citep{Grindlay95}, they can still undergo 
DN-like outbursts though shorter outbursts are then expected 
if they are DQ Her systems. Longer time-based monitoring with HST 
to measure the outburst timescales can distinguish these possibilities. 

\begin{figure}
\begin{center}
{\includegraphics[width=6.7cm]{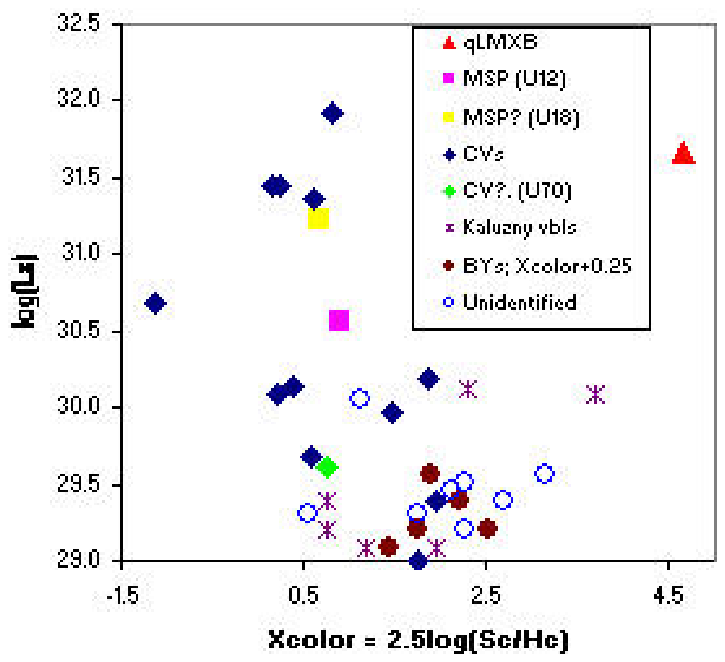}}
{\includegraphics[width=6.7cm]{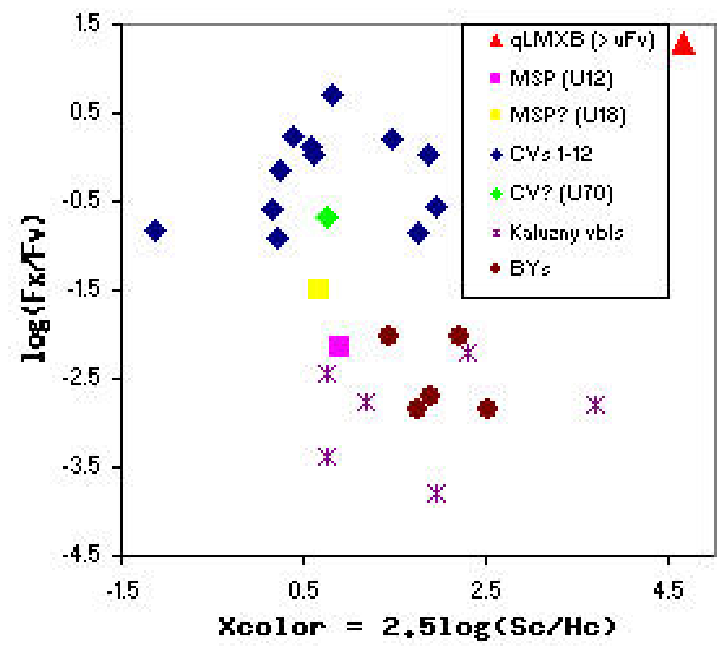}}
\end{center}
\caption{X-ray color-magnitude diagram for 37 sources within the 
cluster half-mass radius plotted for unabsorbed 
luminosity in the Sc band (left) 
and Xcolor (hardness ratio) vs. unabsorbed \FxFv (right) for the 26  
sources in NGC 6397 with optical IDs.} 
\label{fig:XCMD-FxFv-1keV}
\end{figure}

The CVs also stand out from the cluster ABs in their X-ray colors. 
In Figure \ref{fig:XCMD-FxFv-1keV} we plot the X-ray 
color magnitude diagram \citep{Grindlay01a} (XCMD) 
using the hardness ratio Xcolor derived from the ratio of counts in 
the Sc/Hc bands for the 37 sources within the cluster half-mass 
radius. An additional 12 unidentified sources cannot be plotted 
since their total counts are too few to yield meaningful Xcolor 
values. The XCMD is plotted once again in 
the soft (Sc) band, but results 
are very similar in the Hc band. We also show the X-ray 
color vs. \FxFv ratio, which again demonstrates how the CVs 
are separated. We mark a possible CV candidate, U70, since it 
is in the CV domain in the XCMD. It may be identified 
with the He-WD candidate PC-5 reported by \citet{Taylor01}, for 
which the V (F555W) magnitude 22.76 is used to compute \FxFv. The 
\Halpha absorption of the WD masks any \Halpha 
emission. It would thus resemble the lowest \Lx CVs U31 and U60 
and if confirmed would increase the CV total in NGC 6397 to 13. 
Full details of the X-ray spectra, luminosities, 
temporal variability (including the X-ray detection of 
binary periods for several of the CVs) are given by 
GvBB06, and details of new HST identifications 
are given by CLC06. 

Since the \FxFv distribution appears to readily identify CVs, 
we can more confidently compare the CV vs. AB, MSP and qLMXB populations 
in globular clusters than by comparing XCMDs or \Lx distributions 
alone. Additional details are given in GvBB06, but 
a comparison with 47Tuc with the expanded sample of CBs in 
NGC 6397 extends the earlier conclusion of \citet{Grindlay05}: the 
CVs are over-produced relative to ABs in NGC 6397.  The optical 
IDs yield a relative fraction of identified CVs to ABs,  f$_{CV/AB}$ =  
N$_{CV}$/N$_{AB}$ = 12/12 = 1.0$\pm$0.41 in the central portion (Fig. 1) 
of NGC 6397 vs. the ranges for 
N$_{CV}$ vs. N$_{AB}$ derived for 47Tuc by \citet{Heinke05} 
which give f$_{CV/AB}$ = (24-113)/(89-178) = 0.51$\pm$0.38. 

\begin{figure*}
\begin{center}
{\includegraphics[width=12cm]{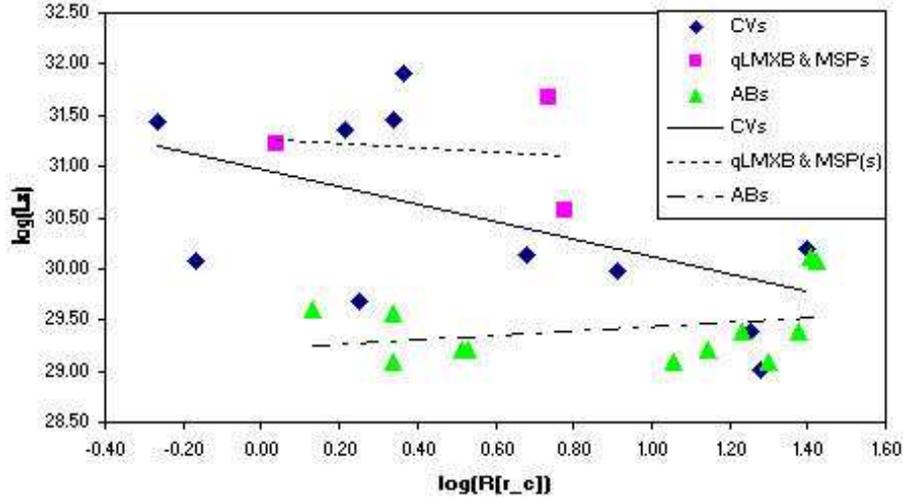}}
\end{center}
\caption{X-ray luminosity (in Sc band) vs. radial offset 
(in units of cluster core radius) for 
sources in NGC 6397. The most luminous CVs are centrally 
concentated, whereas the two lowest luminosity CVs are at 
(nearly) the largest offsets. CVs may be preferentially created during 
core collapse, while ABs are preferentially ``burned'' in the  
central core.}
\label{fig:CVs-NSs-ABs-Lx-radial}
\end{figure*}

It is also instructive to compare the spatial distributions of 
the CBs (CVs vs. ABs, MSPs and qLMXB) in NGC 6397. 
The distribution of the CVs in Figure \ref{fig:merge-I-S-ACS} 
has most of the highest luminous CVs nearest the cluster 
center, surrounded by the ABs. The MSP (U12) and qLMXB (U24) 
are also at larger offsets, though the MSP candidate (U18) is 
in the central concentration of luminous CVs. In 
Figure \ref{fig:CVs-NSs-ABs-Lx-radial} we plot the X-ray luminosity vs. 
radius (in core radius units, adopting r$_c$ = 4.4\arcsec 
from TGE06 for the CVs, NS systems (qLMXB and MSP(s)) 
and ABs together with the regression lines for each. The scatter is 
appreciable but is consistent with a radial dependence of CV 
luminosity (only). The two lowest luminosity CVs, 
U60 (CV9) and U31 (CV11), 
are at the largest offsets. Their optical counterparts (CLC06),  
discovered with new HST/ACS photometry, show them to be 
``quiescent CVs'', with relatively weak \Halpha 
emission -- presumably due to the emission being offset in part 
by the broad \Halpha absorption from the white dwarf. This 
suggests the oldest CVs are farthest from the core; and that 
the luminous systems near the cluster center may then have 
formed in the (most recent) core collapse episode.  As CVs (or 
any CBs) ``age'' they are more likely to be scattered out of 
the core by binary-binary or binary-stellar encounters (which 
may be disruptive). Recently formed (\lsim10$^8$y) CBs, however, 
would be most likely to be filling their Roche lobes and the 
most luminous accretors. Depending on the binary parameters, 
more luminous systems (e.g. U7 and U10; see  Figure \ref{fig:merge-I-S-ACS}) 
can still be maintaining higher mass transfer long 
after having been scattered out of the core 
region in which they were most likely created. 

The fact that the ABs appear (Figure \ref{fig:merge-I-S-ACS}) 
to ``surround'' the CVs in the core (though 
three [U15, U20, and U70] are at comparably small offsets 
from the cluster center) is consistent with binary ``burning'' 
in the central most core during the core collapse episode. 

\section{Other PCC Clusters}
The web-based version of the \citet{Harris96} catalog of globular clusters 
lists 30 globulars (of 148 in the Galaxy with full data) 
as being possible PCC clusters. We drop those that are only 
possibly PCC (listed as ?), leaving 21 as almost certainly PCC. 
Of these, only 4 have been 
observed with Chandra to probe their low luminosity compact binary 
population down to luminosities \Lx \gsim10$^{31}$ \lcgs (Liller 1, 
the Rapid Burster, was also observed but only in a shallow HRC 
observation). Several PCC globulars contain persistently 
luminous LMXBs (e.g. NGC 6624 and M15), which preclude 
even Chandra studies of the low 
luminosity source populations in the cluster cores given the 
bright wings of the psf from the luminous LMXB(s) in the core.  
So, in addition to NGC 6397, only 3 other globulars (Terzan 1, 
NGC 6752 and NGC 7099/M30) are available. Of the 111 globulars that 
are well described by a King model (i.e. isothermal and non-core
collapsed) radial profiles, and excluding again the 
luminous LMXB clusters or those observed with the 
Chandra gratings (e.g. Terzan 2), 11 (47Tuc, NGC 5139, 5272, 6093, 
6121, 6205, 6266, 6440, 6626, 6652, and Terzan 5)  
have been observed with Chandra 
and have published data available. We use the compilation of early 
Chandra results provided by \citet{Heinke03} as well as from 
\citet{Cackett06} for Terzan 1 and from \cite{Lugger06} for 
NGC 7099.

\begin{figure*}
\begin{center}
{\includegraphics[width=6.6cm]{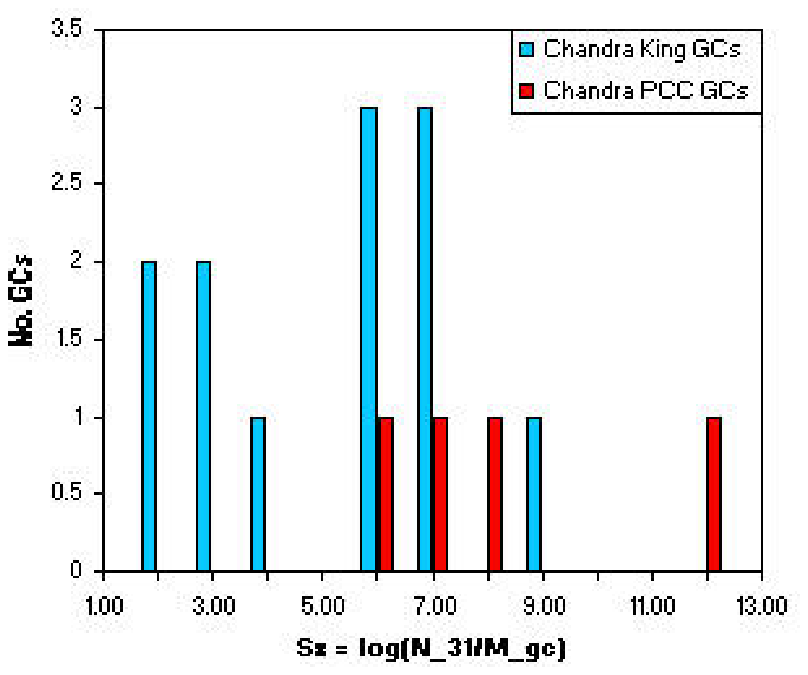}}
{\includegraphics[width=6.5cm]{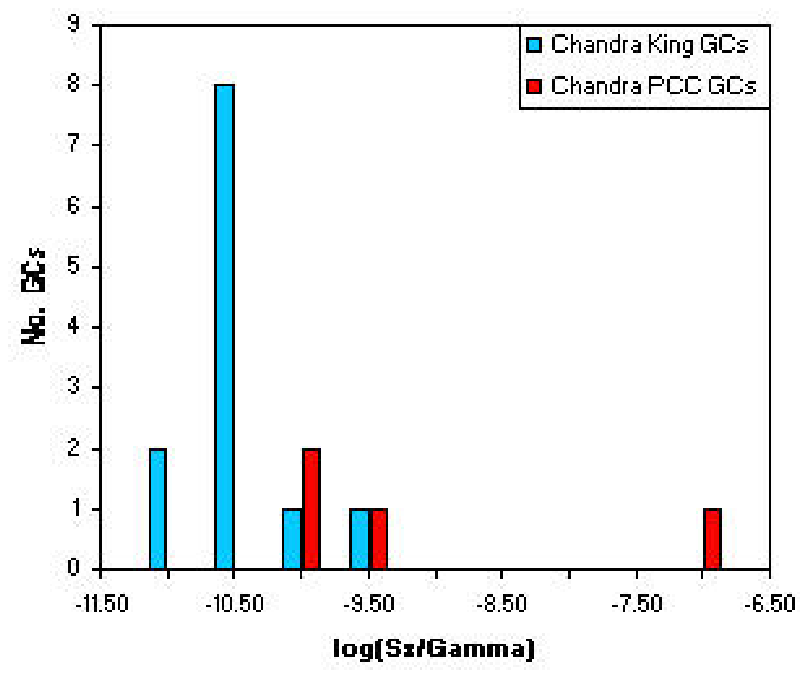}}
\end{center}
\caption{Specific frequency of CBs in GCs for all 16 clusters 
observed with Chandra showing their contributions from 
King model clusters (blue) vs. PCC clusters (red). Distributions 
are plotted for Sx = number of sources per unit cluster mass (left) 
and for Sx per unit collision rate (right).}
\label{fig:SxperGamma}
\end{figure*}

In Figure 
\ref{fig:SxperGamma} we plot the specific frequency of CBs 
per globular, or number of sources per unit cluster mass (left) and 
further normalized for the collision rate $\Gamma \propto 
\rho^{1.5} r_c^2$ in the core (with stellar density $\rho$ and 
core radius r$_c$) of each cluster (right). We derive Sx as 
the number of CBs with \Lx \gsim10$^{31}$ \lcgs~ in order to 
discriminate against chromospherically active binaries (ABs). Thus 
Sx is an approximate measure of the CV, qLMXB and (most luminous) MSP 
content of a globular cluster. We intentionally do not restrict 
the source count to be within just the central core, but rather the 
cluster half-mass radius, in order to include those systems that 
have been scattered out of the core. The larger area included thus 
increases the probability of background source contributions to Sx, 
but this is still small for most clusters. In both Sx 
and particularly for Sx normalized to the stellar encounter rate, 
$\Gamma$, in the cluster core, there is evidence that 
PCC clusters are indeed more abundant in their efficiency of 
production of accretion-powered CBs. 

\section{Discussion}
The spatial distribution, and relative numbers, of CVs vs. ABs 
in NGC 6397 provide direct clues to processes occuring during 
cluster core collapse. The overabundance of CVs vs. ABs in or 
near the cluster center provides the 
the first direct evidence that CBs (including CVs, 
and perhaps qLMXBs) are produced, rather than destroyed, in the 
core collapse process itself. In contrast, Figure \ref{fig:merge-I-S-ACS} 
suggests that ABs, the (usually) less tightly bound ms-ms binaries 
(e.g. BY Dra systems) that are generally surviving primordial 
cluster binaries, are in fact destroyed in the PCC process: no 
ABs (green circles marking source positions) are in the central 
\about10\arcsec of the cluster core. All are in fact {\it exterior} 
to the CVs (red circles). As noted above where ranges are given, the 
numbers of CVs vs. ABs in NGC 6397 appear to be different from what 
is seen in 47Tuc. Using the firm IDs for ABs vs. CVs in 47 Tuc 
\citep{Heinke05} the ratio of ABs:CVs is 
60:22 vs. 12:12 (or possibly 12:13 if U70 is also a CV) in NGC 6397. 
The clear inference is that ABs 
have indeed been ``burned'' (dynamically disrupted) in the 
core collapse process while at the same time, CVs have not been 
destroyed but rather (when scaled per unit cluster mass and 
collision number) in fact created.

Another fact presents itself: the distribution of sources (ABs, CVs, 
qLMXBs and MSPs) in NGC 6397 is decidely {\it anisotropic}. We have commented 
on this for NGC 6397 before \citep{Grindlay01b, Grindlay05}, noting 
that the sources are predominantly to the South East (SE) 
of the cluster center 
and appear in a roughly linear configuration aligned with the 
5 brightest blue stragglers \citep{Grindlay01b} and possibly 
consistent with the plane of the apparent cluster rotation 
equatorial plane \citep{Grindlay05}. With the addition now of the PCC 
cluster Terzan 1 \citep{Cackett06} to the Chandra inventory, it joins 
NGC 6397 and NGC 6752 \citep{Pooley02} in having decidedly 
non-spherical distribution of sources around the cluster center. 
In NGC 6397, at radii \gsim20\arcsec from the cluster center, no 
cluster sources are found to the NE (U28 is a background AGN) 
whereas they extend beyond 
1.9\arcmin~ (U5) to the SE; and in NGC 6752, 13 of the 14 brightest 
sources are S of the cluster center. In Terzan 1, the  13 
brightest sources are aligned in a symmetric (about the cluster center) 
linear configuration (NW-SE). These may all be relatively small N 
statistics chance configurations, and simulations are planned, but 
the similarity is striking and begs explanation. Whereas ``linear'' 
or bar-like configurations may reflect scattering effects 
driven by angular momentum in 
core collapse, the asymmetries in NGC 6397 and NGC 6752 suggest 
proper motion effects (both are roughly aligned; but in the opposite 
sense, when put in the frame of the galactic halo 
(GvBB06)). 

\section{Conclusions}
Compact binaries (CBs) of all stellar types are the engine 
that drive globular clusters away from 
completing their core collapse to form IMBHs.  High resolution 
X-ray images are the most complete, and direct, way to capture 
the CB snapshot since examples of all known CB types are already 
detected or expected. AM CVn's (CVs with WD secondaries) 
should be detected in deeper exposures, or may already 
have been found in NGC 6397 as the lowest \Lx CVs such 
as U60 (CV9), U31 (CV11) and possibly U70. Deep X-ray images thus   
can enable, and test, reconstructions of the dynamical 
interactions in cluster cores. The paradigm that CBs 
are destroyed to halt core collapse (``binary burning'') appears 
borne out in the population of primordial main sequence binaries, 
detected as ABs in X-rays. Perhaps not surprisingly, the CVs 
are the winners: exchanges of isolated cluster WDs into the 
initially much more abundant ABs destroy ABs but produce CVs in the 
cluster core of a PCC cluster. The high CV/AB ratios in NGC 6397 
and NGC 6752 and probably also M30 support this claim. As usual, 
much deeper observations are needed. NGC 6397, as the closest PCC 
cluster, is the most promising prospect to probe the {\it full} CB 
population: both ABs and quiescent CVs can (and will) lurk at 
X-ray luminosities \Lx \lsim5 \X 10$^{29}$ \lcgs, as made clear by the 
discovery of the three new CVs, U25, U60 and U31 with log(L[Sc]) = 
29.69, 29.38 and 29.00, respectively. 

Likewise, and not otherwise 
discussed here, the CBs containing NS primaries (rather than WDs), 
namely the qLMXBs and their eternally-living (and thus dominant 
population) descendents, the MSPs, are another important probe. 
The Chandra source U18 appears in its X-ray spectral properties 
very similar (GvBB06) to the one confirmed MSP (U12) 
and has an optical counterpart that is also similar as a red straggler 
(TGE06; we note that the counterpart suggested by \citet{Kaluzny06} is 
adjacent, but not astrometrically consistent with U18). Both U12 and 
U18 (if confirmed by radio detection) are re-exchanged binaries, again 
indicative of the high interaction rate locally in a PCC core. We also 
note the possibility of a significant number of ``quiescent MSPs'', 
such as the two lowest X-ray luminosity 
MSPs in the field, J1024-0719 and J1744-1134, detected 
with ROSAT \citep{BT99}. Both have X-ray luminosities 
(\Lx \about1 and \about4 \X 10$^{29}$ \lcgs, respectively)  fixed by 
their parallax distances (\lsim200pc and 350pc, 
respectively) and could be detected in NGC 6397 in a deep 
exposure. Whereas both of these field systems are isolated, in PCC 
clusters they should also be found in CBs by exchange collisions. 
Subsequent exchanges (again, particularly in PCC clusters) then 
suggest a population of double NS binaries as the 
graveyard of CBs containing NSs (just as the AM CVn's are 
the corresponding repository of quiescent CVs) in globular clusters. 
Both types may be abundant in PCC globular clusters, since these 
have preferentially exchanged WDs (primarily) and NSs into their 
primordial AB populations and then CB populations 
during what are likely to be repeated core collapse 
oscillations. The final result is 
then that PCC globulars are preferred sites for  both WD-WD mergers 
(and thus possible SNIa events) and for NS-NS mergers 
(and thus a significant fraction of the short Gamma-ray 
Bursts -- as we realized only recently \citep{GPZM06}).

\medskip

{\bf Acknowledgements:} I thank M. van den Berg and S. Bogdanov 
for discussions. This work was supported in part by Chandra grant 
AR6-7010X.



\begin{thebibliography}{}


\bibitem[Bahcall and Wolf(1976)]{Bahcall76}
Bahcall, J.~N. and Wolf, R.~A. Oct. 1976. Star distribution around a 
massive black hole in a globular cluster. Astrophysical Journal, 209, 
214-232.

\bibitem[Becker and Trumper (1999)]{BT99}
Becker, W. and Trumper, J. Jan. 1999. The X-ray emission properties of 
millisecond pulsars. 
Astronomy and Astrophysics, 341, 803-817

\bibitem[Bogdanov et al.(2005)]{Bogdanov05} Bogdanov, S., 
Grindlay, J.~E., \& van den Berg, M.\ Sept. 2005. 
An X-Ray Variable Millisecond Pulsar in the Globular Cluster 
47 Tucanae: Closing the Link to Low-Mass X-Ray Binaries. 
Astrophysical Journal, 630, 1029-1036.

\bibitem[Cackett et al (2006)]{Cackett06} Cackett, E.~M., Wijnands, R., 
Heinke, C. O. et. al.\ 2005. A Chandra X-ray observation of the globular 
cluster Terzan 1. 
Monthly Notices of the Royal Astronomical Society, in press 
(astro-ph/0512168).


\bibitem[Cool \& Bolton(2002)]{Cool02} Cool, A.~M., \& Bolton, 
A.~S.\ 2002. Blue Stars and Binary Stars in NGC 6397: Case Study of a 
Collapsed-Core Globular Cluster. ASP Conf. Series, 263, 163-177. 

\bibitem[Grindlay et al.(1995)]{Grindlay95} Grindlay, J.~E., Cool, 
A.~M., Callanan, P.~J., Bailyn, C.~D., Cohn, H.~N., \& Lugger, P.~M.\ 
Dec. 1995. 
Spectroscopic Identification of Probable Cataclysmic Variables in the Globular 
Cluster NGC 6397. Astrophysical Journal Letters, 455, L47-L51. 
 
\bibitem[Grindlay et al.(2001a)]{Grindlay01a} Grindlay, J. E., Heinke, C., 
Edmonds, P. D., Murray, S. S., Jun. 2001. 
High-Resolution X-ray Imaging of a Globular Cluster Core: Compact Binaries in
47 Tuc. Science, 292, 2290-2295.

\bibitem[Grindlay et al.(2001b)]{Grindlay01b} Grindlay, J.~E., 
Heinke, C.~O., Edmonds, P.~D., Murray, S.~S., \& Cool, A.~M.\ Dec. 2001, 
Astrophysical Journal, 563, L53-L56.
 
\bibitem[Grindlay et al.(2002)]{Grindlay02} Grindlay, J.~E., 
Camilo, F., Heinke, C.~O., Edmonds, P.~D., Cohn, H., \& Lugger, Dec. 2002.  
Chandra Study of a Complete Sample of Millisecond Pulsars in 47 Tucanae 
and NGC 6397. Astrophysical Journal, 581, 470-484. 
 
\bibitem[Grindlay(2005)]{Grindlay05} Grindlay, J.~E.\ 2005. 
Interacting X-ray Binaries in Globular Clusters: 47Tuc vs. NGC 6397. AIP 
Conf.~Proc.~797: Interacting Binaries: Accretion, Evolution, and Outcomes, 
797, 13-22.

\bibitem[Grindlay, Portegies Zwart and McMillan 2006]{GPZM06} 
Grindlay, J., Portegies Zwart, S. and McMillan, S. Feb. 2006, 
Short gamma-ray bursts from binary neutron star mergers in 
globular clusters. Nature Physics, 2, 116-119.


\bibitem[Harris (1996)]{Harris96}Harris, W. E., Oct. 1996. A Catalog of 
Parameters for Globular Clusters in.
the Milky Way. Astronomical Journal 67, 1487-1488.

\bibitem[Heinke et al.(2003)]{Heinke03} Heinke, C. O., 
Grindlay, J. E., Lugger, 
P. M., Cohn, H. N., Edmonds, P. D.,
Lloyd, D. A., Cool, A. M., Nov. 2003. Analysis of the Quiescent Low-Mass
X-Ray Binary Population in Galactic Globular Clusters. Astrophysical
Journal 598, 501-515.

\bibitem[Heinke et al.(2005)]{Heinke05} Heinke, C.~O., Grindlay, 
J.~E., Edmonds, P.~D., Cohn, H.~N., Lugger, P.~M., Camilo, F., Bogdanov, 
S., \& Freire, P.~C. Jun. 2005. A Deep Chandra Survey of the 
Globular Cluster 47 Tucanae: 
Catalog of Point Sources. Astrophysical Journal, 625, 796-824. 

\bibitem[Kaluzny et al. (2006)]{Kaluzny06} Kaluzny, J., Thompson, 
I.~B., Krzeminski, W., Schwarzenberg-Czerny, A.\ Jan. 2006. 
Photometric study of the variable star population in the globular 
cluster NGC 6397. Monthly Notices 
of the Royal Astronomical Society, 365, 548-554.

\bibitem[Lugger et al. (2006)]{Lugger06} Lugger, P.~M., Cohn, H.~C., 
Heinke, C.~O., Grindlay, J.~E., and Edmonds, P.~D. 2006, Chandra 
X-ray Sources in the Collapsed-Core Globular Cluster
M30 (NGC~7099). Astrophysical Journal, in press. 

\bibitem[Pooley et al. (2002)]{Pooley02} Pooley, D., Lewin, W., 
Homer, L., et al. Apr. 2002, Optical Identification of Multiple Faint 
X-Ray Sources in the Globular Cluster NGC 6752: Evidence for 
Numerous Cataclysmic Variables. 
Astrophysical Journal, 569, 405-417.

\bibitem[Shara et al.(2005)]{Shara05} Shara, M.~M., Hinkley, 
S., Zurek, D.~R., Knigge, C., \& Dieball, A.\ Oct. 2005. 
Erupting Cataclysmic Variable Stars in the Nearest Globular Cluster, 
NGC 6397: Intermediate Polars? Astronomical Journal, 130, 1829-1833. 

\bibitem[Taylor et al. (2001)]{Taylor01} Taylor, J.~M., Grindlay, J.~E., 
Edmonds, P.~M., Cool, A.~M. Jun. 2001. Helium White Dwarfs and BY Draconis 
Binaries in the Globular Cluster NGC 6397. Astrophysical Journal, 553, L169


\bibitem[Verbunt et al. (1997)]{Verbunt97} Verbunt, F., Bunk, W.~H., 
Ritter, H., Pfeffermann, E. Nov. 1997. Cataclysmic
variables in the ROSAT PSPC All Sky Survey. Astronomy and Astrophysics
327, 602-613.

\bibitem[Warner(1995)]{w95} Warner, B. 1995, Cataclysmic Variable 
Stars (Cambridge: CUP)

\end{thebibliography}
\end{document}